\documentclass[11pt,preprint]{aastex}
\usepackage{epsfig,lscape}
\slugcomment{KSUPT-03/1 \hspace{0.5truecm} January 2003}

\tighten
 
\begin{document}
\title{Galactic Foreground Constraints from the Python V Cosmic Microwave Background Anisotropy Data}
\author{Pia Mukherjee\altaffilmark{1,2}, Kim Coble\altaffilmark{3,4}, 
        Mark Dragovan\altaffilmark{5}, Ken Ganga\altaffilmark{6},
        John Kovac\altaffilmark{3}, Bharat Ratra\altaffilmark{1}, 
        and Tarun Souradeep\altaffilmark{1,7}}

\altaffiltext{1}{Department of Physics, Kansas State University, 116 Cardwell
                 Hall, Manhattan, KS 66506.}
\altaffiltext{2}{Present address: Department of Physics and Astronomy, 
                 University of Oklahoma, 440 W. Brooks Street, Norman, OK 73019.}
\altaffiltext{3}{Enrico Fermi Institute, University of Chicago, 5640 S. Ellis
                 Ave., Chicago, IL 60637.}
\altaffiltext{4}{Adler Planetarium and Astronomy 
                 Museum, 1300 S. Lake Shore Dr., Chicago, IL 60605.}
\altaffiltext{5}{Jet Propulsion Laboratory, California Institute of 
                 Technology, 4800 Oak Grove Drive, 169-506, Pasadena, CA  
                 91109.}
\altaffiltext{6}{Infrared Processing and Analysis Center, California Institute 
                 of Technology, 100-22, Pasadena, CA 91125.}
\altaffiltext{7}{Permanent address: IUCAA, Post Bag 4, Ganeshkhind, Pune 
                 411007, India.}

\begin{abstract}
We constrain Galactic foreground contamination of the Python V cosmic 
microwave background anisotropy data by cross correlating it with 
foreground contaminant emission templates. To model foreground
emission we use 100 and 12 $\mu$m dust templates and two point source 
templates based on the PMN survey. The analysis takes account of 
inter-modulation correlations in 8 modulations of the data that  
are sensitive to a large range of angular scales and also densely 
sample a large area of sky. As a consequence the 
analysis here is highly constraining. We find little evidence for 
foreground contamination in a analysis of the whole data set. However, 
there is indication that foregrounds are present in the data from the 
larger-angular-scale modulations of those Python V fields that overlap 
the region scanned earlier by the UCSB South Pole 1994 experiment. This 
is an independent consistency cross-check of findings from the South 
Pole 1994 data.
\end{abstract}

\section{Introduction}
While cosmic microwave background (CMB) anisotropy data have started to 
provide interesting constraints on cosmological parameters (see, e.g., 
Podariu et al. 2001; Page 2002; Mukherjee et al. 2002b, 2003; Beno\^{\i}t 
et al. 2003; Ruhl et al. 2002; Kuo et al. 2002; Slosar et al. 2002 for
recent results, and Peebles \& Ratra 2003 for a review), Galactic 
emission foreground contaminants in them are still not well understood. Robust
constraints on cosmological parameters from these data require a better 
understanding of the effect of these contaminants.

In this paper we study foreground contaminants in the Python V (hereafter
PyV) CMB anisotropy data. Python V is the latest of the Python experiments 
at the South Pole. Coble et al. (1999, 2003) describe the PyV experiment, 
observations, and data reduction. Dragovan et al. (1994), Ruhl et al. (1995),
and Platt et al. (1997) describe Python I--III and Rocha et al. (1999) 
and Mukherjee et al. (2003) derive constraints on cosmological parameters
from these data. Coble et al. (2003) also describe the procedure used
to create maps of the sky with PyV and Python III data; these maps were
compared to infer consistency and indirectly deduce the lack of
significant foreground contamination in these data.

The PyV data are acquired at a frequency of $\sim 40$ GHz. Two regions of sky 
covering 598 deg$^2$ in the southern hemisphere were observed (the ``main 
PyV'' region and
a smaller region, the fields labelled `sa', `sb', and `sc' in Coble et al. 
2003, that encompasses the region scanned earlier by the UCSB South Pole 
1994 experiment --- hereafter the ``SP94 overlap'' region). 
690 fields were scanned in all (345 effective fields were scanned
with 2 detector feeds separated by $2.\!\!^\circ 8$ in azimuth 
on the sky) with an asymmetric Gaussian beam of FWHM $0.\!\!^\circ 94 
\times 1.\!\!^\circ 02$. Once the telescope was positioned on each 
field, the chopper smoothly scanned in azimuth with a throw of 
17$^\circ$, 128 samples were recorded in each chopper cycle, and 164 
chopper cycles of data were taken of a given field. The densely sampled 
data were then modulated in software using the first eight cosine harmonics 
of the chopper cycle (hereafter modulations --- or in Tables mods --- 1-8).  
The modulations approach used has the advantage of filtering out some of 
the contaminants in the time stream, and also provides a rapid means of 
compressing a large amount of data into a more manageable size. All the 
modulations, other than the first, were apodized by a Hann window to 
reduce ringing in multipole space and down weight data taken during 
chopper turnaround. The resulting data are sensitive
to angular scales corresponding to multipole moments ranging from $l \sim 40$ 
to $l \sim 260$.  More details about the particular observing strategy 
employed and results found, such as the angular power spectrum of the data, 
may be found in Coble et al. (2003); see Souradeep \& Ratra (2001) for 
details about the window functions. 

For this foregrounds analysis, we follow the general method outlined for
example in Hamilton \& Ganga (2001) and Mukherjee et al. (2002a), doing a 
multi-modulation analysis here rather than the multi-(frequency) channel
analysis discussed in these papers. In doing so we extend the preliminary 
estimate presented in Coble et al. (1999). Here we use the 
technique of Souradeep \& Ratra (2001) to account for correlations between 
modulations, in data that are sensitive to a substantial range of angular 
scales, as well as compare to what has been found from the SP94 
experiment (Gundersen et al. 1995; Ganga et al. 1997) data about
foreground contamination in part of the PyV region (Hamilton \& Ganga 2001; 
Mukherjee et al. 2002a). 

CMB data have previously been correlated with foreground templates (DMR: 
Kogut et al 1996a, 1996b; 19GHz: de Oliveira-Costa et al. 1998; Saskatoon: 
de Oliveira-Costa et al. 1997; OVRO: Leitch et al. 1997, Mukherjee et al. 
2002a; SP94: Hamilton \& Ganga 2001, Mukherjee et al. 2002a; Tenerife: 
de Oliveira-Costa et al. 1999, 2002, Mukherjee et al. 2001; QMAP: 
de Oliveira-Costa et al. 2000; MAX: Lim et al. 1996, Ganga et al.
1998; and Boomerang: Masi et al. 2001)\footnote{
DASI (Halverson et al. 2002) and VSA (Taylor et al. 2002) are 
interferometric experiments, that observed at frequencies of 
26-36 GHz and are sensitive to multipoles of $\sim$ 100 to 900. 
Point sources are the dominant source of contamination and the 
effect of contamination from diffuse Galactic foregrounds upon the 
CMB power spectrum is inferred to be small for both data sets. A 
full cross-correlation analysis of VSA data is underway (Dickinson 
et al., in preparation).}.  
Correlations between CMB data and infra-red emission seem to be roughly 
consistent with free-free emission, spectrally, over a wide range of 
frequencies (10 to 90 GHz) and angular scales ($7^\circ$ to $7^\prime$), 
with some evidence for a contribution from spinning dust emission with a 
peak around 15 to 20 GHz.  In general though the contamination from
 foregrounds has not been found to be large 
in any experiment, but on the detail level residual foregrounds can 
cause problems with parameter estimation and non-Gaussianity tests for high 
precision CMB data.\footnote{The CMB anisotropy is often thought to have 
been generated 
by quantum mechanical fluctuations in a weakly coupled scalar field during
an early epoch of inflation and thus would be a realization of a spatially
stationary Gaussian random process (see, e.g., Ratra 1985; Fischler, Ratra,
\& Susskind 1985). Measurements lend fairly strong support to this
Gaussianity assumption (see, e.g., Park et al. 2001; Shandarin et al. 2002;
Santos et al. 2002; Polenta et al. 2002). See Park, Park, \& Ratra (2002)
for the effects of foreground contamination on Gaussianity tests based
on anticipated MAP data.} 

To model diffuse Galactic emission, we use the 100 $\mu$m IRAS+DIRBE map 
(Schlegel, Finkbei\-ner, \& Davis 1998) as a tracer of thermal emission 
from interstellar dust, and the 12 $\mu$m map (D. Finkbeiner, private 
communication, 2000) as a tracer of emission from ultra-small dust grains.  
Emission from small dust grains is still under study (Finkbeiner et al. 
2002). Such grains may contribute significantly at microwave frequencies 
according to a model by Draine \& Lazarian (1998a,b), and this has been 
reviewed from the CMB data point of view by Kogut (1999) and Draine \& 
Lazarian (1999). The Haslam 408 MHz map (Haslam et al. 1981) of synchrotron 
emission was not used as it does not have enough resolution for all the 
modulations of the PyV experiment to be simulated on it. At the same time 
synchrotron emission is not likely to contribute significantly at 40 GHz. 
We have tried to use the SHASSA $H_\alpha$ data (Gaustad et al. 2001) as a 
tracer of free-free emission, but found that the template has insufficient 
resolution for our purpose.  

We also use two point source templates created from the PMN survey (Wright 
et al. 1994). The one called PMN has been converted to $\delta T_{CMB}$ 
(equivalent temperature fluctuations in the CMB at 40 GHz) using the spectral 
indices given in the survey, while the one called PMN0 is converted to a flux 
at 40 GHz assuming a flat spectrum extrapolated from the flux measurement at 
4.85 GHz.  The assumption of a flat spectrum is conservative in that it is 
likely to overestimate the flux at 40 GHz. Neither case is correct, since 
spectral indices have not been measured for all of the sources, in which case 
a flat spectrum is assumed. The two cases cover a reasonable range of 
possibilities.  These templates were also used in the preliminary foreground 
analysis of Coble et al. (1999).

Each of the Galactic emission maps have been converted into a template for cross correlation with the PyV data by simulating the PyV observing strategy on it, taking account of the asymmetric beam.   
Since a chopper synchronous offset and a ground shield offset were removed 
from the data, we account for this by adding the chopper synchronous offset 
and ground shield constraint matrices to the noise matrix and marginalizing 
over them in the analysis.  The CMB theory covariance matrix is modelled 
using a spatially-flat cosmological constant dominated CDM model with 
non-relativistic matter density parameter $\Omega_0 = 0.3$, scaled baryonic
matter density parameter $\Omega_B h^2 = 0.021$ (here $h$ is the present
value of the Hubble parameter in units of 100 km s$^{-1}$ Mpc$^{-1}$), and 
age $t_0=14$ Gyr (Ratra et al. 1999a) with the CMB anisotropy normalized 
to a quadrupole moment amplitude $Q_{\rm rms-PS} = 20$ $\mu$K. 

\section{Correlations}

\subsection{Method}

We follow the general method outlined for example in Hamilton \& Ganga (2001)
and Mukherjee et al. (2002a), doing a multi-modulation analysis here rather
than the multi-(frequency) channel analysis discussed in these papers.
The method assumes that the data are a linear combination of CMB 
anisotropies and foreground components, 
\begin{equation}
   y = aX + x_{\rm CMB} + n.
\end{equation} 
Here $y$ is a 
$n_{\rm mod} \times N$ element vector containing the data, $n$ is the 
corresponding noise vector, $x_{\rm CMB}$ is the CMB signal, and $X$ is a 
$(n_{\rm mod} \times N) \times n_{\rm mod}$ element matrix 
containing the simulated foreground template, $N$ being the number of data 
points per modulation and $n_{\rm mod}$ being the number of modulations.
In a given column $X$ contains mostly zeros except for the rows 
corresponding to that modulation, where it contains the simulated template 
signal. The vector $a$ contains $n_{\rm mod}$ elements that represent the 
amplitude of the correlated foreground signal. If the noise and CMB 
anisotropies are uncorrelated Gaussian distributed variables
with zero mean, minimizing $\chi^2$ leads to a best fit estimate of 
\begin{equation}
  \hat{a} = [X^{T} C^{-1} X]^{-1} X^{T} C^{-1} y,
\end{equation} 
where $C$ is the $(n_{\rm mod} \times N) \times (n_{\rm mod} \times N)$ 
element total covariance matrix (sum of the theory covariance matrix 
that models the CMB signal, the noise covariance matrix, and any constraint 
matrices). The vector $\hat{a}$ then contains the best estimate for the 
correlation slopes for the corresponding template for all modulations, 
given all information about inter-modulation correlations. The matrix 
\begin{equation}
  \Sigma = \langle \hat{a}^2 \rangle - \langle \hat{a} \rangle^2 = 
  [X^{T} C^{-1} X]^{-1}, 
\end{equation}
is its covariance matrix. The rms amplitudes of temperature fluctuations 
in the data that results from the correlation is $\Delta T = 
(\hat{a}\pm \delta\hat{a}) \sigma_{\rm{fore}}$, where $\sigma_{\rm{fore}}$ 
contains the rms deviations of the corresponding foreground template in 
the different modulations.

The method assumes that our foreground emission maps are good enough to
model the foregrounds in the data accurately in all the different 
modulations, at a frequency different from that of the original 
foreground emission map, and that eq. (1) explains all the structure 
in all modulations of the data. If we further assume that the ratio of 
the signal in the data and that in the foreground template is the same for 
all the modulations, then a net correlation slope can be found using
\begin{equation}
  \bar{a}=\frac{\rm{Total}(\Sigma^{-1}a)}{\rm{Total}(\Sigma^{-1})}.
\end{equation} 
Here Total denotes the sum of all elements of a matrix or vector.  This 
is thus a weighted average taking account of correlations between the $a$
values of the different modulations. We may choose not to treat all 
modulations the same (see discussion in $\S$ 2.3). The error bar on $\bar{a}$
is $\sqrt{({\rm Total}(\Sigma^{-1}))^{-1}}$.

\subsection{Results}

The correlation slopes obtained from a complete inter-modulation analysis, 
using all the PyV data points, for each foreground emission template 
individually, are given in Table 1 (top panel). We do not find any 
significant correlation. The net correlation slopes (i.e., the weighted mean 
of the correlation slopes given in the table, taking cross-modulation 
correlations into account) are not significant, with two standard deviation 
upper limits of 34 $\mu$K(MJy/sr)$^{-1}$, 110 $\mu$K(MJy/sr)$^{-1}$, 
156 $\mu$K/$\mu$K, and 897 $\mu$K(MJy/sr)$^{-1}$, for the 100 $\mu$m, the 
12 $\mu$m, the PMN, and the PMN0 templates, respectively. These limits on 
100 $\mu$m and PMN0 correlation slopes compare well with those of Table 2 
of Coble et al. (1999).  (There are some errors in the PMN correlation slopes 
of Table 2 and the rms values in Table 3 of Coble et al. 1999.)  
The upper limits obtained from using all modulations of the data together 
are given in Table 3.

The results of repeating this analysis for just the SP94 overlap region 
are also given in Table 1 (bottom panel).\footnote{
An outlier that affects 6 data points per feed per modulation is seen in the 
12 $\mu$m template, hence these data points are ignored whenever this region 
is being analysed. Outliers are harder to pick out over larger regions of sky 
but they also affect the result less, so nothing is removed when analysing the 
full data set.}  
The uncertainties in the estimated correlation slopes are higher here, with 
the number of data points down by a factor of almost 8 (the time spent 
observing this region was less than 1:8 of the total observing time), 
but the level of 
associated temperature fluctuations are relatively higher, at least for the 
100 $\mu$m template.  As seen from Tables 2 and 3, the rms of the data and 
the 100 $\mu$m template are higher in this patch of sky. Again, no significant 
correlations are found. The weighted correlation coefficient between
 foreground emission templates is given by
 $\Sigma_{ij}^{-1} (\Sigma_{ii}^{-1} \Sigma_{jj}^{-1}) ^{-0.5}$ 
(de Oliveira-Costa et al. 1999). The 100 and 12 $\mu$m templates and 
the PMN and PMN0 templates are found to have significant weighted 
correlation coefficients in this region of sky, with the correlation 
reducing somewhat, from 0.7 to 0.5, with increasing modulation for the
two dust templates, while it remains steady at about 0.85 for the
 two point source templates.
(These 
templates are much less like each other when all the PyV fields are 
considered.) Yet when all modulations of the data are analysed together, 
joint template fits do not detect any correlations. The net correlation 
slopes obtained from a complete inter-modulation analysis of the SP94 
overlap region are not significant, with 2 $\sigma$ upper limits of 
69 $\mu$K(MJy/sr)$^{-1}$, $1027$ $\mu$K(MJy/sr)$^{-1}$, $603$ $\mu$K/$\mu$K, 
and $2198$ $\mu$K(MJy/sr)$^{-1}$, for the 100 $\mu$m, the 12 $\mu$m, the PMN, 
and the PMN0 templates, respectively. 
The upper limits obtained from using all modulations of the data together 
are given in Table 3.

\subsection{Low $l$ Modulations}

As seen from Table 2, the signal in the foreground templates falls 
more steeply with increasing modulation number than does the data. So the 
case here, of fitting for several modulations simultaneously, is somewhat 
different from the case of fitting multi-frequency data: if a certain 
foreground is present in the data, picking it out in the higher modulations 
will be harder. But the cumulative effect of the other modulations is 
significant on the estimated correlation slope for any modulation; 
Figure 2 of Coble et al. (1999) shows how much the CMB signal in different 
modulations overlap. Hence simultaneously fitting for the correlation slopes 
in all the modulations of the data may not be the most appropriate thing to 
do. It is also important to note that the foregrounds may not have been 
accurately modelled in the higher modulations because of insufficient 
resolution (given the dense sampling in the data). There may also be 
different kinds of unmodelled or incorrectly modelled foregrounds/errors 
in different modulations. Hence it is useful to also look at the results 
from analysing say the first three modulations together, because across these 
modulations the data and the foreground signal seem to be roughly similar 
as regards rms values, or each modulation individually, at the cost of 
increased uncertainty in the estimates.

If we look at only the first modulation in the SP94 overlap region, we find 
a 1.8 $\sigma$ correlation slope of $80 \pm 45$ $\mu$K(MJy/sr)$^{-1}$ 
($36 \pm 21$ $\mu$K) for the 100 $\mu$m template (Table 4). For the 
12 $\mu$m template the correlation found is just less than a sigma, but here 
the two dust templates have a weighted correlation coefficient of 0.65, and 
performing a joint correlation of the data with these two templates results 
in a 1.9 $\sigma$ correlation slope of $115 \pm 60$ $\mu$K(MJy/sr)$^{-1}$ 
($52 \pm 27$ $\mu$K) for the 100 $\mu$m template at the cost of a 1 $\sigma$ 
negative correlation with the 12 $\mu$m template. This may be the signal 
found in the foregrounds analysis of the SP94 data (Hamilton \& Ganga 2001; 
Mukherjee et al. 2002a; for earlier qualitative indications see Ganga et al.
1997; Ratra et al. 1999b). While there a detection was aided by the presence 
of 7 frequency channels, here the error bars for the individual modulation 
correlations are large. This could be just a chance correlation on the other 
hand, and the probability of that is reflected in the significance of the 
result. It might also be relevant to note here that the SP94 data was only 
one dimensional while here the PyV `sa', `sb', and `sc' data are essentially
two dimensional.

If we look at only the second modulation by itself, the correlation slopes are
as shown in Table 4, and even though the correlation coefficient between the 12
and 100 $\mu$m templates is 0.70 nothing is gained by a joint fit this time.
 
In the first two modulations together some correlation with the 100 $\mu$m 
template is detected consistent with the above estimates. When the first three 
modulations are correlated simultaneously, a 1.6 $\sigma$ correlation slope of 
$58\pm36$ $\mu$K(MJy/sr)$^{-1}$ ($25\pm16$ $\mu$K) is found with the 
100 $\mu$m template in the first modulation and a 1 $\sigma$ $710\pm695$ 
$\mu$K(MJy/sr)$^{-1}$ ($14\pm14$ $\mu$K) correlation is found with the 12 
$\mu$m template in the second modulation. The correlation coefficient between 
the templates is high and correlations of similar significance are found in 
the same modulations when the two templates are analysed jointly.

Regarding the point source templates, nothing significant shows up in an 
individual template analysis (Table 4). When analysing the first two 
modulations together the two point source templates have a correlation 
coefficient of 0.8 and a correlation slope of $6555 \pm 4994$ 
$\mu$K(MJy/sr)$^{-1}$ ($28 \pm 21$ $\mu$K) at 1.3 $\sigma$ is found with the 
PMN0 template in modulation 1 at the cost of a 1 $\sigma$ negative correlation 
with the PMN template in the same modulation. When analysing the first 
three modulations together the point source templates have a high correlation 
coefficient and a 1.2 $\sigma$ correlation of $5625 \pm 4498$ 
$\mu$K(MJy/sr)$^{-1}$ ($24 \pm 19$ $\mu$K) is found with the PMN0 template 
in modulation 1 at the cost of a $-1$ $\sigma$ correlation with the PMN 
template in the same modulation. 

When all the PyV fields are taken together (690 in each modulation), a 
1 $\sigma$ correlation with the 100 $\mu$m template shows up in the 
first modulation (when the first modulation is analysed by itself, or 
jointly with the second, or jointly with the second and third), and a 
1 $\sigma$ correlation with the PMN0 template shows up in the first 
modulation when it is analysed together with the second modulation and 
jointly with the PMN template.  The uncertainty in the correlation slopes 
with the point source templates is lowest in the third and fourth 
modulations, but nothing significant shows up when these modulations 
are analysed separately. Hence consistent correlations (in the low-$l$ modulations and at low significance) are seen even when both the PyV regions 
are analysed together. 

\subsection{Summary}

When all the modulations of the data are analysed together, no 
significant correlations are found, in the whole PyV data set or in the 
SP94 overlap region. These results are summarized in Table 3 and plotted 
in Figure 1. 

We have discussed the motivations for looking at individual modulations 
separately, and we find some indication of foregrounds in the low 
$l$ modulations. Correlation slopes for some of these low $l$ 
modulations are summarized in Table 4. We see that greater than 
1 $\sigma$ correlations show up in the SP94 overlap region with the 
100 $\mu$m template in particular, but also with the 12 $\mu$m 
template and the PMN0 template in some modulations in a joint 
modulation analysis of a few modulations. And consistent correlations 
are seen when all the PyV data fields are analysed together.

We note that according to 
the foreground model discussed for example in Mukherjee et al. (2002a), if 
the 100 $\mu$m correlations  are from both free-free emission and spinning 
dust emission, then by 40 GHz the contribution from spinning dust emission 
(as traced by the 12 $\mu$m template) may have again dropped below that from
free-free (the frequency at which spinning dust emission peaks depends on 
the details of the spinning dust emission model and this can also vary 
from region to region), so that we would expect 100 $\mu$m correlations to be 
more significant than 12 $\mu$m correlations at this frequency (see Figure 
1$a$). However, given the error bars (the numbers in Table 4), we do not 
expect to see significant correlations in the PyV data (see Figure 1). 


\section{Conclusion}

Using the method of cross correlating CMB anisotropy data with foreground 
contaminant emission templates, we find little evidence for 
foreground contamination in the whole PyV region when analysing 
all modulations of the data together. There is however 
indication of foreground 
contamination in the low-$l$ modulations of the PyV fields that cover the 
region scanned earlier by the SP94 experiment, consistent with results from 
the SP94 data. This is a valuable test, indicating consistency between results
found using data from two different experiments. Given the uncertainties, our 
findings are not inconsistent with the two component dust-correlated 
(free-free and spinning dust) foreground emission model that other data 
sets tentatively seem to point to (Mukherjee et al. 2002a).  

\bigskip

We thank D. Finkbeiner for the 12 $\mu$m dust map and E. Boyce, R. Ekers, 
E. Ryan-Weber, and L. Stavely-Smith for their work on this project. PM, 
BR, and TS acknowledge support from NSF CAREER grant AST-9875031. KC is supported 
by NSF grant AST-0104465. This work was partially carried out at the 
Infrared Processing and Analysis Center and the Jet Propulsion Laboratory 
of the California Institute of Technology, under a contract with the 
National Aeronautics and Space Administration.

\clearpage

\begin{landscape}
\begin{deluxetable}{lccccccccc}
\tablecolumns{9} 
\tablewidth{0pt}
\tablecaption{Results of Correlating the PyV Data with Individual Foreground Emission Templates\tablenotemark{a}}
\tablehead{
\colhead{Template} & \colhead{mod1} & \colhead{mod2} & \colhead{mod3} & \colhead{mod4} & \colhead{mod5} & \colhead{mod6} & \colhead{mod7} & \colhead{mod8} } 
\startdata
100 $\mu$m & $1\pm20$ & $-4\pm20$ & $-20\pm20$ & $-6\pm23$ & $5\pm30$ & $8\pm52$ & $-34\pm110$ & $-172\pm245$ \\
           & $0\pm4$ & $-1\pm3$ & $-2\pm2$ & $0\pm2$ & $0\pm1$ & $0\pm1$ & $0\pm1$ & $-1\pm1$ \\
12 $\mu$m & $-82\pm131$ & $-75\pm123$ & $-79\pm120$ & $-133\pm126$ & $-257\pm140$ & $-372\pm182$ & $-228\pm259$ & $-476\pm516$ \\
          & $-2\pm3$ & $-2\pm3$ & $-1\pm2$ & $-2\pm2$ & $-3\pm1$ & $-3\pm1$ & $-1\pm1$ & $-1\pm1$ \\
PMN & $137\pm123$ & $66\pm108$ & $-14\pm103$ & $-68\pm100$ & $-158\pm105$ & $-173\pm126$ & $-137\pm177$ & $8\pm252$ \\
    & $3\pm3$ & $2\pm3$ & $0\pm2$ & $-1\pm2$ & $-2\pm1$ & $-2\pm1$ & $-1\pm1$ & $0\pm1$ \\
PMN0 & $684\pm502$ & $546\pm408$ & $237\pm382$ & $15\pm369$ & $-111\pm385$ & $-116\pm452$ & $-44\pm625$ & $972\pm986$ \\
     & $3\pm2$ & $3\pm2$ & $1\pm2$ & $0\pm1$ & $0\pm1$ & $0\pm1$ & $0\pm1$ & $1\pm1$ \\ \hline
100 $\mu$m & $19\pm30$ & $18\pm30$ & $5\pm31$ & $12\pm35$ & $30\pm50$ & $92\pm94$ & $-104\pm266$ & $792\pm1051$ \\
           & $8\pm14$ & $6\pm11$ & $1\pm7$ & $2\pm6$ & $3\pm5$ & $4\pm4$ & $-2\pm4$ & $4\pm5$ \\
12 $\mu$m & $116\pm570$ & $235\pm559$ & $-288\pm561$ & $-248\pm647$ & $-348\pm868$ & $-626\pm 1672$ & $-2476\pm3871$ & $7207\pm11467$ \\
          & $2\pm12$ & $4\pm10$ & $-4\pm7$ & $-2\pm6$ & $-2\pm6$ & $-2\pm5$ & $-3\pm5$ & $4\pm6$ \\
PMN & $205\pm327$ & $98\pm291$ & $70\pm266$ & $5\pm276$ & $78\pm320$ & $-123\pm436$ & $319\pm801$ & $862\pm1717$ \\
    & $5\pm8$ & $3\pm8$ & $1\pm6$ & $0\pm4$ & $1\pm4$ & $-1\pm3$ & $1\pm4$ & $3\pm5$ \\
PMN0 & $1871\pm2030$ & $947\pm1722$ & $636\pm1607$ & $267\pm1645$ & $409\pm1958$ & $-1496\pm2851$ & $3495\pm5538$ & $9599\pm11042$ \\
     & $8\pm8$ & $5\pm8$ & $2\pm6$ & $1\pm5$ & $1\pm4$ & $-2\pm4$ & $3\pm4$ & $5\pm6$ \\
\enddata
\tablenotetext{a}{For each template the correlation slopes (in units of 
$\mu$K(MJy/sr)$^{-1}$ for the dust and the PMN0 templates and in units of $\mu$K/$\mu$K for the PMN template) are given in the first row and the corresponding temperature fluctuations (in $\mu$K) are given in the second row.  In the top panel we have used all the data points, and in the bottom panel we use only the fields in the SP94 overlap region.  Cross-modulation correlations are accounted for and the 8 modulations represent decreasing angular scales covering $l \sim 40$ to $\sim 260$.}
\end{deluxetable}
\end{landscape} 

\begin{deluxetable}{lcccccc}
\tablecolumns{6} 
\tablewidth{0pt}
\tablecaption{Rms Values of the Signal in the PyV Data and the Templates{\tablenotemark{a}}}
\tablehead{
\colhead{Modulation} & \colhead{Data} & \colhead{100 $\mu$m} & \colhead{12 $\mu$m} & \colhead{PMN} & \colhead{PMN0} \\
 & $\mu$K & MJy/sr & MJy/sr & $\mu$K & MJy/sr }
\startdata
1 & 92.0 & 0.222 & 0.022 & 0.022 & 0.004 \\
2 & 95.1 & 0.170 & 0.022 & 0.027 & 0.006 \\
3 & 79.4 & 0.100 & 0.019 & 0.022 & 0.005 \\
4 & 77.0 & 0.078 & 0.015 & 0.017 & 0.004 \\
5 & 78.6 & 0.047 & 0.010 & 0.013 & 0.003 \\
6 & 77.4 & 0.023 & 0.007 & 0.010 & 0.002  \\
7 & 69.8 & 0.011 & 0.006 & 0.007 & 0.002 \\
8 & 65.0 & 0.006 & 0.003 & 0.005 & 0.001 \\ 
\hline
1 & 123.0 & 0.455 & 0.022 & 0.025 & 0.004 \\
2 & 129.9 & 0.356 & 0.018 & 0.028 & 0.005 \\
3 & 104.9 & 0.221 & 0.013 & 0.022 & 0.004 \\
4 & 111.1 & 0.180 & 0.010 & 0.016 & 0.003 \\
5 & 113.6 & 0.103 & 0.007 & 0.012 & 0.002 \\
6 & 111.5 & 0.042 & 0.003 & 0.008 & 0.001  \\
7 & 95.7 & 0.015 & 0.001 & 0.004 & 0.001 \\
8 & 89.33 & 0.005 & 0.001 & 0.003 & 0.0005 \\
\enddata
\tablenotetext{a}{The top panel is for the whole PyV region with 690 fields per modulation. The bottom panel is for the SP94 overlap region with 90 fields per modulation.}
\end{deluxetable}

\begin{deluxetable}{lcccccc}
\tablecolumns{6} 
\tablewidth{0pt}
\tablecaption{Correlation Slopes}
\tablehead{
\colhead{Fields} & \colhead{Modulations} & \colhead{100 $\mu$m} & 
\colhead{12 $\mu$m} & \colhead{PMN} & \colhead{PMN0} \\
 & & $\mu$K(MJy/sr)$^{-1}$ & $\mu$K(MJy/sr)$^{-1}$ & $\mu$K/$\mu$K & $\mu$K(MJy/sr)$^{-1}$ }
\startdata
all & all & $-4\pm19$ & $-120\pm115$ & $-30\pm93$ & $209\pm344$ \\
\vspace{3mm}
SP94 overlap & all & $13\pm28$ & $-21\pm524$ & $85\pm259$ & $904\pm1551$ \\
\enddata
\end{deluxetable}

\begin{deluxetable}{lcccccc}
\tablecolumns{6} 
\tablewidth{0pt}
\tablecaption{Correlation Slopes for Low $l$ Modulations{\tablenotemark{a}}}
\tablehead{
\colhead{Fields} & \colhead{Modulations} & \colhead{100 $\mu$m} & 
\colhead{12 $\mu$m} & \colhead{PMN} & \colhead{PMN0} \\
 & & $\mu$K(MJy/sr)$^{-1}$ & $\mu$K(MJy/sr)$^{-1}$ & $\mu$K/$\mu$K & $\mu$K(MJy/sr)$^{-1}$ }
\startdata
all & 1st 3 mods & $7\pm22$ & $-155\pm141$ & $-59\pm131$ & $203\pm553$ \\
all & 1st 2 mods & $16\pm24$ & $-70\pm162$ & $-77\pm154$ & $416\pm658$ \\
all & 1st mod & ${\bf{29}}\pm{\bf{29}}$ & $97\pm223$ & $-161\pm213$ & $213\pm903$ \\
all & 2nd mod & ${\bf{29}}\pm{\bf{33}}$ & $-27\pm202$ & $-53\pm190$ & $271\pm779$ \\ \hline
SP94 overlap & 1st 3 mods & ${\bf{33}}\pm{\bf{32}}$ & $388\pm590$ & $-129\pm332$ & $-283\pm1924$ \\
SP94 overlap & 1st 2 mods & ${\bf{50}}\pm{\bf{36}}$ & $522\pm674$ & $-6\pm387$ & $736\pm2215$  \\
SP94 overlap & 1st mod & ${\bf{80}}\pm{\bf{45}}$ & $406\pm875$ & $190\pm563$ & $2131\pm3401$ \\
SP94 overlap & 2nd mod & ${\bf{49}}\pm{\bf{47}}$ & $854\pm889$ & $65\pm473$ & $-158\pm2805$ \\
\enddata
\tablenotetext{a}{Boldface type indicates 1 $\sigma$ detection.}
\end{deluxetable}

\begin{figure}
\centerline{\epsfig{file=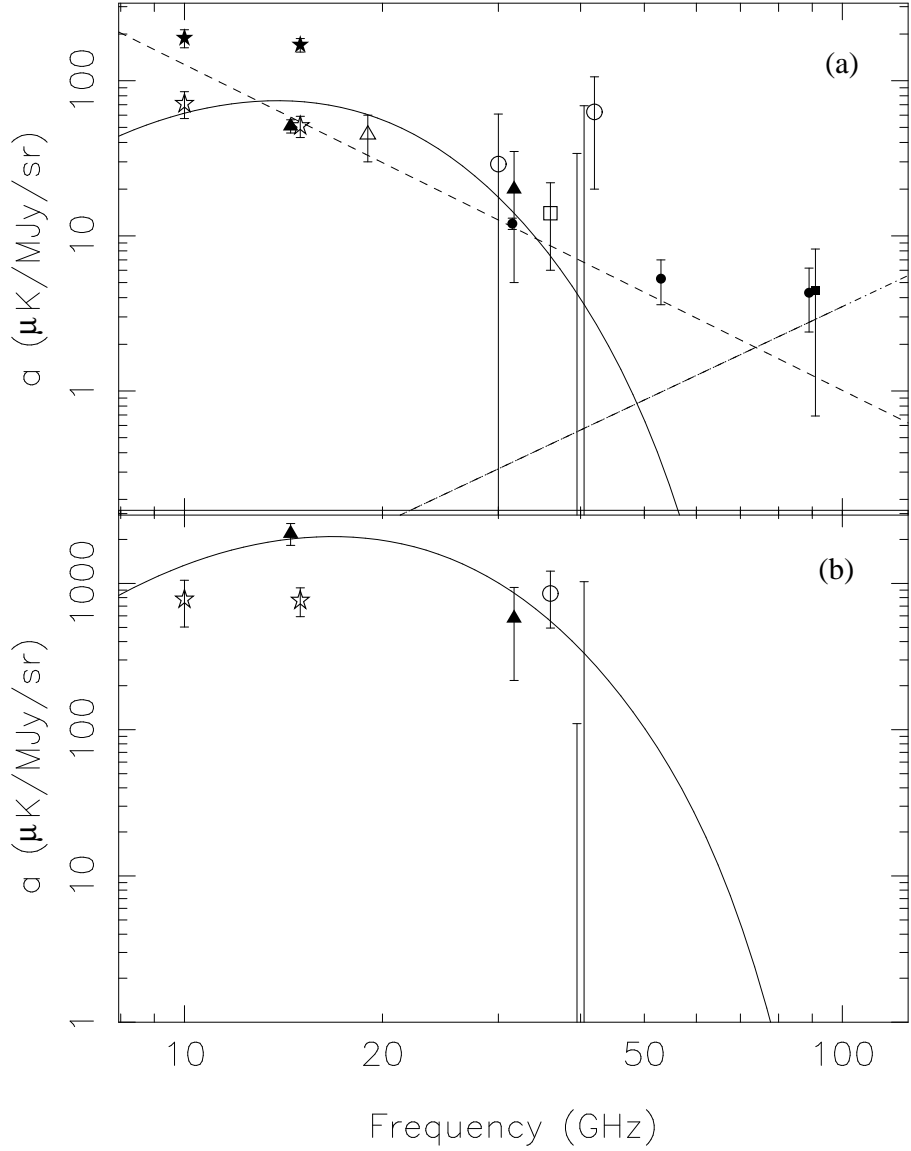,width=12cm}}
\caption{$a)$ Correlation slopes between CMB anisotropy data and the dust 
100 $\mu$m template. Two standard deviation upper limits obtained for PyV 
from analysing all the fields and from analysing the SP94 overlap region 
only are both shown at $\sim40$ GHz. $b)$ Correlations with the 12 $\mu$m 
template. Results are also shown for other experiments: Tenerife (open stars, 
de Oliveira-Costa et al. 2001; filled stars, Mukherjee et al. 2001), 
OVRO (filled triangles), 19GHz (open triangles), SP94 (open circles), 
Saskatoon (open square), Boomerang (filled rectangle), and DMR (filled 
circles).  The solid curves are representative of a spinning dust spectrum, 
the dashed and the dot-dashed lines represent free-free and vibrational 
dust emission spectra, respectively.}
\label{sdplot}
\end{figure} 


\clearpage

\begin{thebibliography}{99}

\bibitem[Beno\^{\i}t et al. (2003)]{benoit}Beno\^{\i}t, A., et al. 2003, A{\&}A, 399, L25

\bibitem[Coble et al. (1999)]{coble1999}Coble, K., et al. 1999, ApJ, 519, L5

\bibitem[Coble et al. (2003)]{coble2003}Coble, K., Dodelson, S., Dragovan, M.,
  Ganga, K., Knox, L., Kovac, J., Ratra, B., \& Souradeep, T. 2003, \apj, 
  584, 585

\bibitem[de Oliveira-Costa et al. (1997)]{costa97}
  de Oliveira-Costa, A., Kogut, A., Devlin, M. J., Netterfield, C. B., 
  Page, L. A., \& Wollack, E. J. 1997, ApJ, 482, L17

\bibitem[de Oliveira-Costa et al. (2000)]{costa00}de Oliveira-Costa, A., et al. 2000, ApJ, 542, L5

\bibitem[de Oliveira-Costa et al. (2002)]{costa02}de Oliveira-Costa, A., et al. 2002, ApJ, 567, 363

\bibitem[de Oliveira-Costa et al. (1999)]{costa99}de Oliveira-Costa, A., 
  Tegmark, M., Gutierrez, C. M., Jones, A. W., Davies, R. D., Lasenby, A. N., 
  Rebolo, R., \& Watson, R. A. 1999, ApJ, 527, L9

\bibitem[de Oliveira-Costa et al. (1998)]{costa98}de Oliveira-Costa, A., 
  Tegmark, M., Page, L. A., \& Boughn, S. P. 1998, ApJ, 509, L9

\bibitem[Dragovan et al. 1994]{dragovan94} Dragovan, M., Ruhl, J. E., 
  Novak, G., Platt, S. R., Crone, B., Pernic, R., \&\ Peterson, J. B. 
  1994 ApJ, 427, L67

\bibitem[Draine \& Lazarian (1998a)]{draine98a}
  Draine, B. T., \& Lazarian, A. 1998a, ApJ, 494, L19

\bibitem[Draine \& Lazarian (1998b)]{draine98b}
  Draine, B. T., \& Lazarian, A. 1998b, ApJ, 508, 157

\bibitem[Draine \& Lazarian (1999)]{draine99}
  Draine, B. T., \& Lazarian, A. 1999, in ASP Conf. Ser. 181, Microwave 
  Foregrounds, ed. A. de Oliveira-Costa \& M. Tegmark (San Francisco: ASP), 133

\bibitem[Finkbeiner et al. (2002)]{finkbeiner}Finkbeiner, D. P., Schlegel, 
  D. J., Frank, C., \& Heiles, C. 2002, ApJ, 566, 898

\bibitem[Fischler et al. (1985)]{fischler85}
  Fischler, W., Ratra, B., \& Susskind, L. 1985, Nucl. Phys. B, 259, 730

\bibitem[Ganga et al. (1997)]{ganga97}
  Ganga, K., Ratra, B., Gundersen, J. O., \& Sugiyama, N. 1997, ApJ, 484, 7

\bibitem[Ganga et al. (1998)]{ganga98}Ganga, K., Ratra, B., Lim, M. A.,
  Sugiyama, N., \& Tanaka, S. T. 1998, ApJS, 114, 165

\bibitem[Gaustad et al. (2001)]{gaustad01}Gaustad, J. E., McCullough, P. R., 
  Rosing, W., \& Van Buren, D. 2001, PASP, 113, 1326

\bibitem[Gundersen et al. (1995)]{gundersen95}
  Gundersen, J. O., et al. 1995, ApJ, 443, L57

\bibitem[Halverson et al. (2002)]{halverson2002}
  Halverson, N. W., et al. 2002, ApJ, 568, 38

\bibitem[Hamilton, \& Ganga (2001)]{hamilton}Hamilton, J.-Ch., \& Ganga K.M. 2001, A\&A, 368, 760

\bibitem[Haslam et al. (1981)]{haslam81}
  Haslam, C. G. T., Klein, U., Salter, C. J., Stoffel, H., Wilson, W. E., 
  Cleary, M. N., Cooke, D. J., \& Thomasson, P. 1981, A{\&}A, 
  100, 209 

\bibitem[Kogut (1999)]{kogut99}Kogut, A. 1999, in ASP Conf. Ser. 181, Microwave Foregrounds, ed. A. de Oliveira-Costa \& M. Tegmark (San Francisco: ASP), 91

\bibitem[Kogut et al. (1996a)]{kogut96a}Kogut, A., Banday, A. J., Bennett, 
  C. L., G\'orski, K. M., Hinshaw, G., \& Reach, W. T. 1996a, ApJ, 460, 1

\bibitem[Kogut et al. (1996b)]{kogut96b}Kogut, A., Banday, A. J., Bennett, 
  C. L., G\'orski, K. M., Hinshaw, G., Smoot, G. F., \& Wright, E. L. 1996b, 
  ApJ, 464, L5

\bibitem[Kuo et al. (2002)]{kuo}Kuo, C. L., et al. 2002, astro-ph/0212289

\bibitem[Leitch et al. (1997)]{leitch97}Leitch, E. M., Readhead, A. C. S., 
    Pearson, T. J., \& Myers, S. T. 1997, ApJ, 486, L23

\bibitem[Lim et al. (1996)]{lim}Lim, M. A., et al. 1996, ApJ, 469, L69 

\bibitem[Masi et al. (2001)]{masi}
Masi, S., et al. 2001, ApJ, 553, 93 

\bibitem[Mukherjee et al. (2002a)]{mkherjee02a}
  Mukherjee, P., Dennison, B., Ratra, B., Simonetti, J. H., Ganga, K., \& 
  Hamilton, J.-Ch. 2002a, ApJ, 579, 83

\bibitem[Mukherjee et al. (2003)]{mukherjee03}
  Mukherjee, P., Ganga, K., Ratra, B., Rocha, G., Souradeep, T., 
  Sugiyama, N., \& G\'orski, K. M. 2003, Intl. J. Mod. Phys., in press, 
  astro-ph/0209567

\bibitem[Mukherjee et al. (2001)]{mukherjee01}
  Mukherjee, P., Jones, A. W., Kneissl, R., \& Lasenby, A. N. 2001, MNRAS, 
  320, 224

\bibitem[Mukherjee et al. (2002b)]{mukherjee02b}
  Mukherjee, P., Souradeep, T., Ratra, B., Sugiyama, N., \& G\'orski, K. M. 
  2002b, astro-ph/0208216

\bibitem[Page (2002)]{page02} 
  Page, L. 2002, astro-ph/0202145

\bibitem[Park et al. (2002)]{park02} Park, C.-G., Park, C., \& Ratra, B.
  2002, ApJ, 568, 9

\bibitem[Park et al. (2001)]{park01} Park, C.-G., Park, C., Ratra, B., \& 
  Tegmark, M. 2001, ApJ, 556, 582

\bibitem[Peebles \& Ratra (2003)]{peebles03} 
  Peebles, P. J. E., \&\ Ratra, B. 2003, Rev. Mod. Phys, in press, 
  astro-ph/0207347

\bibitem[Platt et al. 1997]{platt97} Platt, S. R., Kovac, J., Dragovan, M., 
  Peterson, J. B., \& Ruhl, J. E. 1997, ApJ, 475, L1 

\bibitem[Podariu et al. (2001)]{podariu01} Podariu, S., Souradeep, T., Gott, 
  J. R., Ratra, B., \& Vogeley, M. S. 2001, ApJ, 559, 9

\bibitem[Polenta et al. (2002)]{polenta02}
  Polenta, G., et al. 2002, ApJ, 572, L27

\bibitem[Ratra (1985)]{ratra85}
  Ratra, B. 1985, Phys. Rev. D, 31, 1931

\bibitem[Ratra et al. (1999a)]{ratra99a} Ratra, B., Ganga, K., Stompor, R., 
  Sugiyama, N., de Bernardis, P., \&\ G\'orski, K. M. 1999a, ApJ, 510, 11

\bibitem[Ratra et al. (1999b)]{ratra99b} Ratra, B., Stompor, R., Ganga, K., 
  Rocha, G., Sugiyama, N., \&\ G\'orski, K. M. 1999b, ApJ, 517, 549 

\bibitem[Rocha et al. (1999)]{rocha99} Rocha, G., Stompor, R., Ganga, K.,  
  Ratra, B., Platt, S. R., Sugiyama, N., \& G\'orski, K. M. 1999, ApJ, 525, 1

\bibitem[Ruhl et al. (2002)]{ruhl02}
  Ruhl, J. E., et al. 2002, astro-ph/0212229

\bibitem[Ruhl et al. 1995]{ruhl95} Ruhl, J. E., Dragovan, M., Platt, S. R.,
  Kovac, J., \&\ Novak, G. 1995, ApJ, 453, L1

\bibitem[Santos et al. (2002)]{santos02}
  Santos, M. G., et al. 2002, Phys. Rev. Lett., 88, 241302

\bibitem[Schlegel, Finkbeiner, \& Davis (1998)]{schlegel98}
  Schlegel, D. J., Finkbeiner, D. P., \& Davis, M. 1998, ApJ, 500, 525

\bibitem[Shandarin et al. (2002)]{shandarin02}
  Shandarin, S. F., Feldman, H. A., Xu, Y., \& Tegmark, M. 2002, ApJS, 141, 1

\bibitem[Slosar et al. (2002)]{slosar02}
  Slosar, A., et al. 2002, astro-ph/0212497

\bibitem[Souradeep \& Ratra (2001)]{souradeep01} Souradeep, T., \& Ratra,
  B. 2001, \apj, 560, 28

\bibitem[Taylor et al. (2002)]{taylor2002}
  Taylor, A. C., et al. 2002, MNRAS, submitted, astro-ph/0205381

\bibitem[Wright et al. (1994)]{wright94}
  Wright, A. E., Griffith, M. R., Burke, B. F., \& Ekers, R. D. 1994, ApJS, 
  91, 111

\end{thebibliography}
\end{document}